\documentclass[%
reprint,
superscriptaddress,
showpacs,
amsmath,amssymb,
aps,
prb,
]{revtex4-1}

\usepackage{graphicx}
\usepackage{dcolumn}
\usepackage{bm}

\usepackage{color}

\usepackage{ulem} 

\newcommand{\eval}[3]{\langle#1\vert#2\vert#3\rangle}

\begin{document}


\title{Photoinduced in-gap excitations in the one-dimensional extended Hubbard model}

\author{Hantao Lu}
\email{luht@lzu.edu.cn}
\affiliation{Center for Interdisciplinary Studies $\&$ Key Laboratory for Magnetism and Magnetic Materials of the MoE, Lanzhou University, Lanzhou 730000, China}

\author{Can Shao}
\affiliation{Center for Interdisciplinary Studies $\&$ Key Laboratory for Magnetism and Magnetic Materials of the MoE, Lanzhou University, Lanzhou 730000, China}

\author{Janez Bon\v{c}a}
\affiliation{Faculty of Mathematics and Physics, University of Ljubljana, SI-1000 Ljubljana, Slovenia}
\affiliation{J. Stefan Institute, SI-1000 Ljubljana, Slovenia}

\author{Dirk Manske}
\affiliation{Max-Planck-Institut f\"{u}r Festk\"{o}rperforschung, Heisenbergstrasse 1, D-70569 Stuttgart}

\author{Takami Tohyama}
\email{tohyama@rs.tus.ac.jp}
\affiliation{Department of Applied Physics, Tokyo University of Science, Tokyo 125-8585, Japan}


\date{\today}
     

\begin{abstract}

We investigate the time evolution of optical conductivity in the half-filled one-dimensional extended Hubbard model driven by a transient laser pulse, by using the time-dependent Lanczos method. Photoinduced in-gap excitations exhibit a qualitatively different structure in the spin-density wave (SDW) in comparison to the charge-density-wave (CDW) phase. In the SDW, the origin of a low-energy in-gap excitation is attributed to the even-odd parity of the photoexcited states, while in the CDW an in-gap state is due to confined photogenerated carriers. The signature of the in-gap excitations can be identified as a characteristic oscillation in the time evolution of physical quantities.

\end{abstract}
\maketitle


\section{Introduction}

Time-resolved spectroscopy can provide new insights into the dynamical properties of condensed matter. Among many time-resolved spectroscopies, pump-probe optical measurements have been performed extensively to investigate strongly correlated electron systems. Pumping pulse applied to one-dimensional (1D) Mott insulators with spin-density-wave (SDW) correlation creates not only a photoinduced metallic state~\cite{Okamoto2007,Wall2011} but also a photoinduced charge-density-wave (CDW) state~\cite{Matsuzaki2014}. In two dimensions, the photoexcitation of underdoped cuprates induces a signature of enhanced superconductivity~\cite{Fausti2011,Nicoletti2014,Hu2014}. Even in a conventional superconductor, an observation of the amplitude mode~\cite{Matsunaga2013} is under intensive investigation within the field of nonequilibrium and nonlinear phenomena induced by a transient electric field.

The time evolution of optical conductivity in the ultrafast process has been theoretically examined~\cite{Wall2011,Papenkort2007,Unterhinninghofen2008,Eckstein2008,Kanamori2010,Eckstein2010,Filippis2012,Iyoda2014,Lenarcic2014}, since this quantity contains information on the photoinduced charge dynamics and the initial relaxation through internal degrees of freedom of a given closed system. In Mott insulators, photoexcitation is expected to induce a new excitation channel inside the Mott gap. The nature of those states has not been fully understood from a theoretical viewpoint even for 1D Mott insulators where, due to spin charge separation, the spin degrees of freedom affect the charge dynamics much less than in two-dimensional (2D) Mott insulators~\cite{Okamoto2011,Lenarcic2013}.

In this paper we address the physical properties of photoinduced states, their positions within the Mott gap, and their influence on the time evolution of the nonequilibrium optical properties. For this purpose, we calculate the time-dependent optical conductivity of a 1D extended Hubbard model driven by a transient laser pulse. In this model, one can change the ground state from the SDW phase to the CDW phase by changing the interaction parameters and tuning the laser pulse so as to enhance charge correlation even within the SDW phase~\cite{Lu2012}. We find that photoinduced in-gap excitations appear in a different manner, depending on the nature of the initial ground state, i.e., SDW or CDW state. In the SDW, a dominant in-gap excitation emerges at the low-energy region reflecting the energy difference between the even-odd parity of photoexcited states. In the CDW, in-gap states appear in the middle of the Mott gap and their origin can be attributed to photogenerated carriers. We also find that the signature of the in-gap excitations emerges in the time evolution of physical quantities such as the kinetic energy and peak intensity of the in-gap excitations as a characteristic oscillation.

The rest of the paper is organized as follows. In Sec. II, the model is introduced and the numerical method motivated by the experimental pump-probe setup is described. In Sec. III the characteristic optical responses of the system driven by a laser pulse are presented. The photoinduced in-gap excitations, which exhibit qualitatively different structures in the SDW and CDW phases, are analyzed in detail. Their influence on the time evolution of various physical observables is discussed. Finally, a summary is given in Sec. IV.

\section{Model and Method}

We investigate the extended one-dimensional (1D) Hubbard model at half filling,
\begin{eqnarray}
H&=&-t_h\sum_{i,\sigma}\left(c^{\dagger}_{i,\sigma} c_{i+1,\sigma}+\text{H.c.}\right)+U\sum_{i}\left(n_{i,\uparrow}-\frac{1}{2}\right)\nonumber\\
&&\times\left(n_{i,\downarrow}-\frac{1}{2}\right)
+V\sum_{i}\left(n_{i}-1\right)\left(n_{i+1}-1\right),
\label{H}
\end{eqnarray} 
where $c^{\dagger}_{i,\sigma}$ ($c_{i,\sigma}$) is the creation (annihilation) operator of an electron with spin $\sigma$ at site $i$, $n_i= n_{i,\uparrow}+ n_{i,\downarrow}$, $t_h$ is the hopping constant, and $U$ and $V$ are on-site and nearest-neighbor repulsions, respectively. 

The application of a spatially uniform and time-dependent electric field expressed by a vector potential $A(t)$ is incorporated into the hopping terms via the Peierls substitution:
\begin{equation}
c^{\dagger}_{i,\sigma}c_{i+1,\sigma}+\text{H.c.}\rightarrow
e^{iA(t)}c^{\dagger}_{i,\sigma}c_{i+1,\sigma}+\text{H.c.},
\end{equation}
leading to $H\rightarrow H(t)$. We assume that $A(t)$, representing the pumping pulse, is expressed in terms of the temporal gauge
\begin{equation}
A_\text{pump}(t)=A_0e^{-\left(t-t_0\right)^2/2t_d^2}\cos\left[\omega_0\left(t-t_0\right)\right],
\label{eq:vpotent}
\end{equation}
where the temporal distribution of $A_\text{pump}(t)$ is Gaussian
centered around $t_0$, while the width is controlled by the parameter
$t_d$, and $\omega_0$ is the central frequency. The average incoming
number of photons per lattice site during the pump is estimated to be $\propto A_0^2\omega_0 t_d$.

In our calculation, the system's temporal evolution is evaluated by the time-dependent Lanczos method, where, within the reach of the exact diagonalization, the time-dependent wave function $\psi(t)$ can be obtained exactly~\cite{Prelovsek2013}. Based on the information of $\psi(t)$, we calculate the time-dependent current density $j(t)=\eval{\psi(t)}{\hat{j}}{\psi(t)}$, where $\hat{j}$ is the current operator. 

Motivated by the experimental pump-probe setup, we use the following strategy to obtain the time-dependent optical conductivity after pumping. For a given pump, we trace the temporal evolution of the system twice: first, without the probe and the second time with the  applied probe. The current induced by the probing pulse $j_{\text{pr}}(t)$ is obtained from the current difference between these two consecutive simulations. The subsequent Fourier transformations on $j_{\text{pr}}(t)$ and $A_{\text{pr}}(t)$ produce the time-dependent optical conductivity~\cite{Papenkort2007} as $\sigma\left(\omega,\Delta t\right) \approx j_{\text{pr}}(z) /iz A_{\text{pr}}(z)$, where $\Delta t$ is the time difference between the probing pulse and the pumping pulse and $z=\omega+i\delta$ with a small positive number $\delta$. Introduction of $\delta$ is necessary for obtaining the spectral weight that is due to the Drude peak in the limit of $\omega\rightarrow 0$. We have checked that $\sigma\left(\omega,\Delta t\right)$ gives a good agreement with that obtained by a nonequilibrium linear-response theory~\cite{Lenarcic2014} as long as the strength and the width of the probing pulse are small enough. We note that the result is also insensitive to the details of the probing pulse, regardless of the shape of the vector potential and the central frequency.

In our numerical calculations, we consider a half-filled periodic lattice with $L$ sites. We set $t_h=1$ as a unit of energy and $\hbar=1$. Throughout this work, the value of $U$ is chosen to be $U=10$, which is in the strong coupling regime. The central frequency of the pumping pulse $\omega_0$ is always set to match the absorption-peak energy of a given system. The small positive number $\delta$ for the calculation of $\sigma(\omega,\Delta t)$ is taken to be $\delta=1/L$.

\section{Results and Discussion}
\subsection{In-gap excitations in SDW}

\begin{figure}
\includegraphics[width=0.45\textwidth]{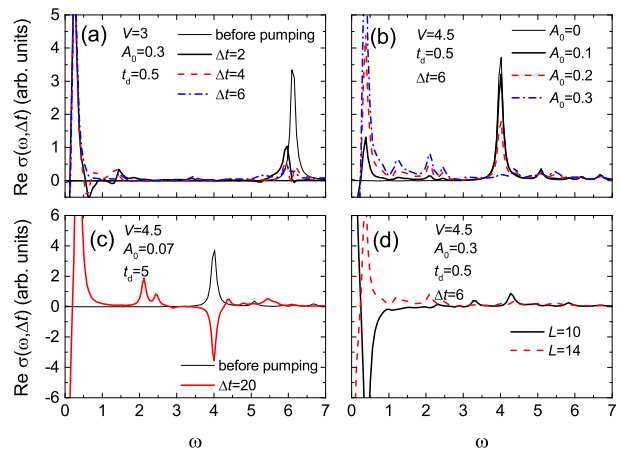}
\caption{(Color online)
$\text{Re}\,\sigma(\omega,\Delta t)$ for the SDW phase of the extended Hubbard model with $U=10$. (a) $\Delta t$ dependence for $L=14$ and $V=3$. $A_0=0.3$ and $t_d=0.5$. (b) $A_0$ dependence for $L=14$ and $V=4.5$. $\Delta t=6$ and $t_d=0.5$. (c) $L=14$, $V=4.5$, $A_0=0.07$, and $\Delta t=20$. $t_d=5$ is ten times larger than the case of (b), resulting in an enhancement of charge correlation after pumping~\cite{Lu2012}. (d) Comparison between $L=10$ and $L=14$ with $V=4.5$, $A_0=0.3$, $\Delta t=6$, and $t_d=0.5$.
}
\label{fig1}
\end{figure}

Let us start with the case where the initial ground state is the SDW state. Figure~\ref{fig1}(a) shows the real part of the time-dependent optical conductivity, $\text{Re}\,\sigma(\omega,\Delta t)$, for a $L=14$ ring with $V=3$. Before pumping, there is a single absorption peak at $\omega=6.1$, which is due to the exciton. With pumping, the exciton absorption peak loses its weight, while a low-energy peak at $\omega=0.2$ emerges in the Mott gap. The low-energy peak persists throughout the time evolution. Figure~\ref{fig1}(b) shows $A_0$ dependence at $\Delta t=6$ for $V=4.5$. As expected, with increasing $A_0$, the exciton peak at $\omega=4$ loses its weight continuously and the low-energy in-gap peak increases.

We now discuss the origin of the low-energy peak seen in Figs.~\ref{fig1}(a) and \ref{fig1}(b). Since the peak appears only after excitation with the pump pulse, its origin should be related to photoexcited states. The photoexcitation excites carriers into the state that gives rise to the absorption peak in the equilibrium, which is an optically allowed odd-parity state. The probe pulse couples in part to the photoinduced carriers in the odd-parity state, resulting in an excitation from the optically allowed state to an optically forbidden even-parity state. In Mott insulators, it is known that the optically forbidden excitonic state is located slightly above the optically allowed excitonic state~\cite{Mizuno2000,Takahashi2002}.  Therefore, nonequilibrium optical conductivity should detect such an excitation. This is the origin of the low-energy peak. Its energy thus corresponds to the energy separation between the optically allowed state (exciton peak) and the forbidden state. In order to confirm this, we perform the same calculation on a smaller system size. The optically forbidden state is known to be lower in energy than the allowed one if the small system has periodic boundary conditions. A $L=10$ lattice with $V=4.5$ is such a case. If the optically forbidden state is lower in energy than the optically allowed one, a negative-energy excitation appears after pumping, leading to negative weight at the positive-energy side since $\omega \sigma(\omega)$ is odd in $\omega$. This actually occurs in the $L=10$ lattice as shown in Fig.~\ref{fig1}(d), where $\text{Re}\,\sigma(\omega,\Delta t=6)$ is compared with that of the $L=14$ lattice. The data for $L=10$ exhibit a peak at $\omega=0.4$ with a negative weight in contrast to the $L=14$ case with a positive weight. This confirms our hypothesis that low-energy in-gap excitation comes from the excitation from the optically allowed to forbidden state.

\begin{figure}
\includegraphics[width=0.25\textwidth, angle=-90]{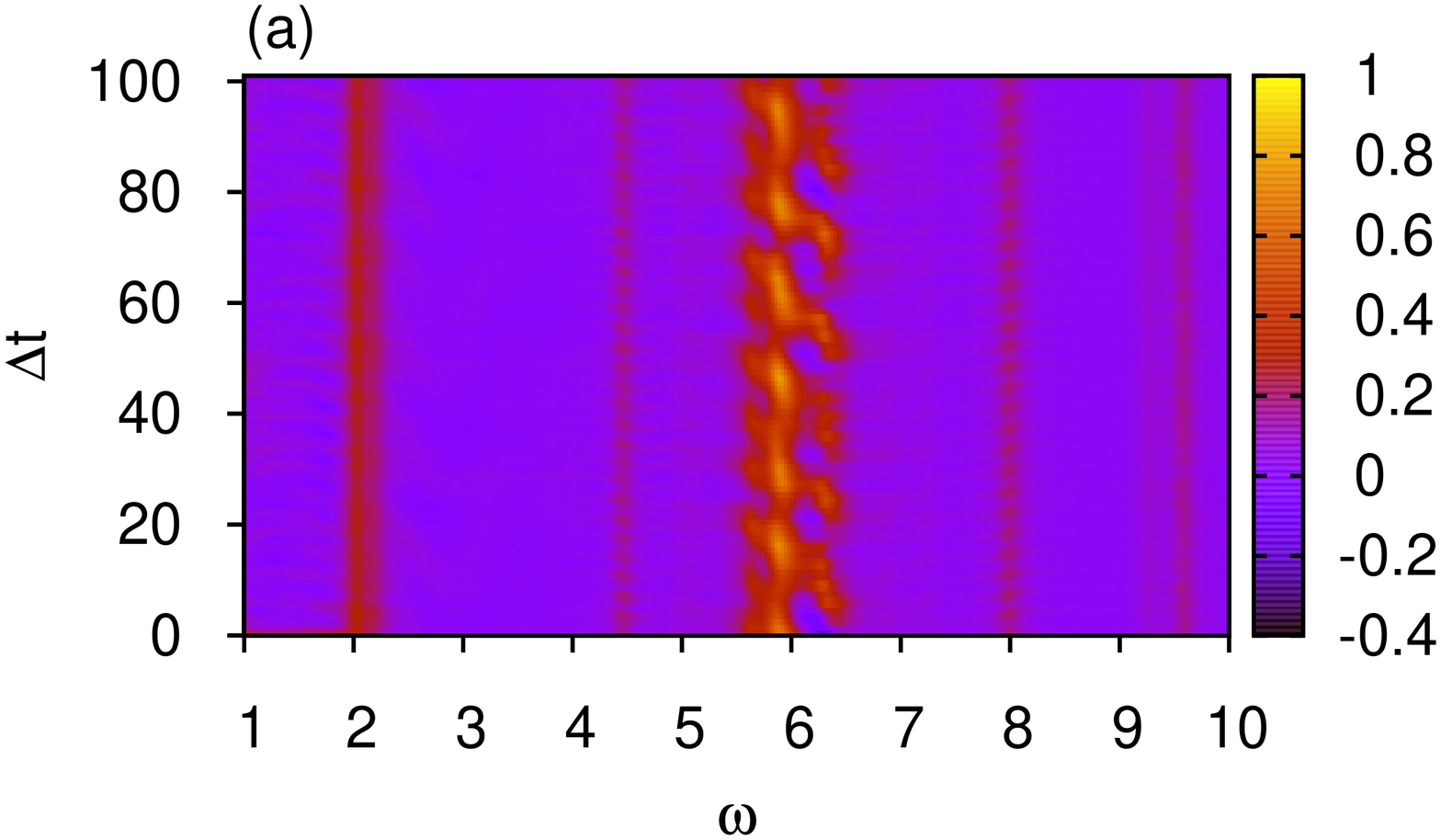}
\includegraphics[width=0.4\textwidth]{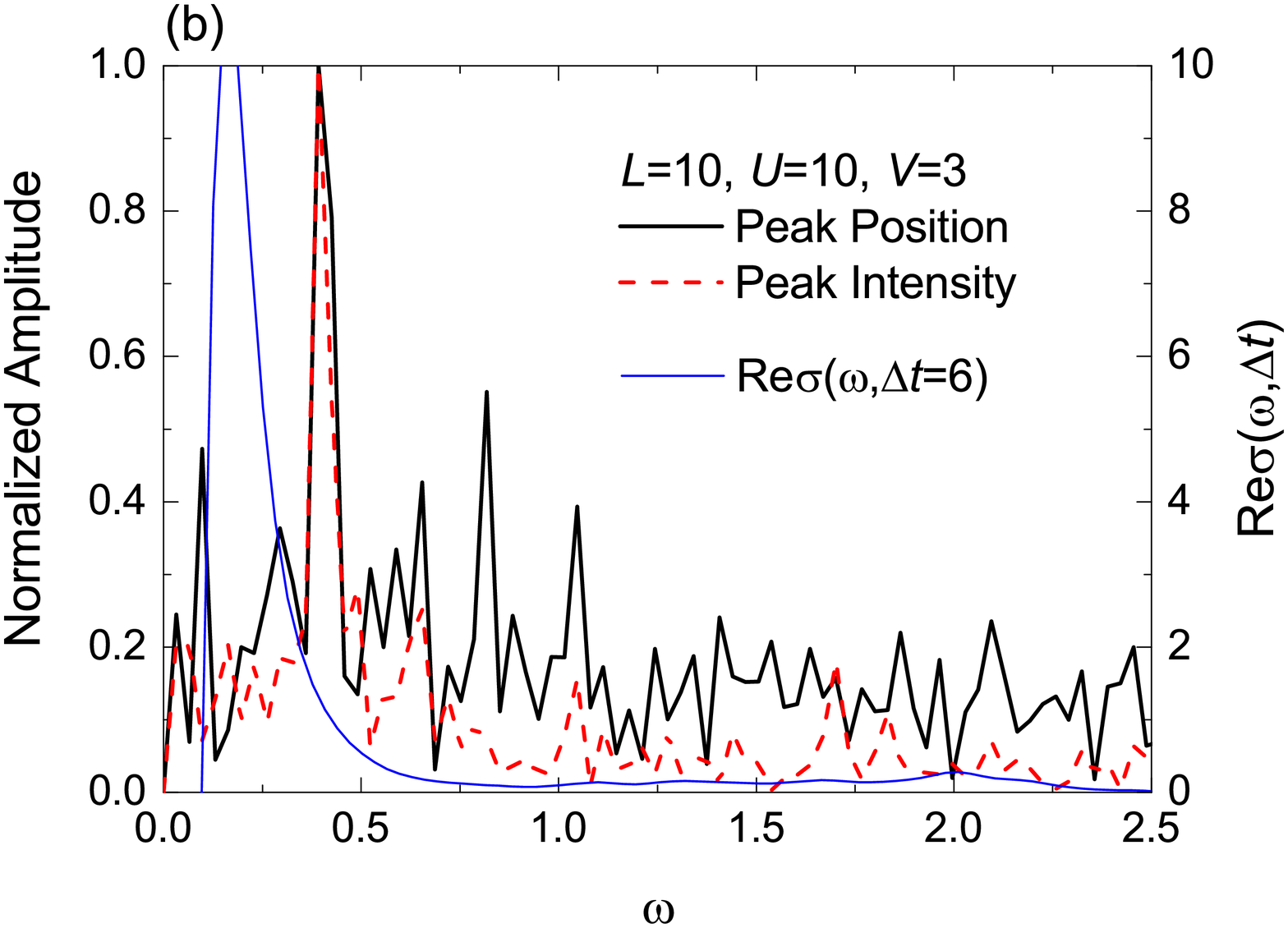}
\caption{(Color online)
(a) Contour plot of $\text{Re}\,\sigma(\omega,\Delta t)$ ($1\le\omega\le10$) for the SDW phase of the $L=10$ extended Hubbard model with $U=10$ and $V=3$. The amplitude is normalized by the maximum value of the equilibrium absorption peak. The conditions of the pumping pulse are $A_0=0.3$ and $t_d=0.5$. (b) The amplitudes of Fourier components of the maximum intensity (red dashed line) near $\omega=6$ and the corresponding position (black solid line) in (a). The amplitude is normalized by the maximum value. The blue thin line represents $\text{Re}\,\sigma(\omega,\Delta t=6)$ for the same system.
}
\label{fig2}
\end{figure}

It is vital to test whether the photoinduced low-energy in-gap excitation demonstrated in Figs.~\ref{fig1}(a) and \ref{fig1}(b) can be detected by pump-probe optical experiments. A recent time-dependent optical reflectivity measurement on {\it bis}(ethylendithyo)-tetrathiafulvalene-difluorotetracyanoquinodimethane ET-F$_2$TCNQ has indicated the presence of photoinduced low-energy states located around $\sim$100~meV by analyzing the intensity oscillation near an exciton peak~\cite{Wall2011}. This energy is well below the energy of the exciton peak, 675~meV. The situation looks similar to Figs.~\ref{fig1}(a) and (b). In order to explore the possibility of detecting the in-gap excitation, we calculate $\text{Re}\,\sigma(\omega,\Delta t)$ and make a Fourier analysis, as is the case of the experiment. Figure~\ref{fig2}(a) shows the contour plot of $\text{Re}\,\sigma(\omega,\Delta t)$ ($1\le\omega\le10$) for a $L=10$ ring with $U=10$ and $V=3$. We note that in this system and parameter set the photoinduced in-gap excitation emerges with a positive-spectral weight at $\omega=0.2$. We can find an oscillating behavior of a photoinduced state below the exciton peak at $\omega=6.3$ as a function of $\Delta t$. The oscillation has a period of $\Delta t\sim 16$. The Fourier amplitudes of the maximum intensity near $\omega=6$ and the corresponding position are shown in Fig.~\ref{fig2}(b), together with $\text{Re}\,\sigma(\omega,\Delta t=6)$. We observe a dominant peak at $\omega\sim 2\pi/16 \sim 0.4$ in both the position and the intensity. The peak position of the Fourier amplitude roughly matches the position of the photoinduced in-gap excitation at $\omega=0.2$, which indicates it might come from the Rabi oscillation of the two odd-even states. We also find that such a low-frequency oscillation close to the energy of the in-gap excitation does not occur in the cases with the negative-weight peak, e.g., $V=2$ at $L=10$ (not shown). Therefore, we consider that the oscillation appearing in the exciton peak reflects the energy of the in-gap excitation.

We next consider the case of a large pump width $t_d=5$ for $V=4.5$, where a strong enhancement of short-range charge correlation after pumping has been reported in Ref.~\onlinecite{Lu2012}. The large width of the pump pulse $t_d$ leads to a well-tuned energy transfer between the pump pulse and the system. After the pump pulse we thus expect an enhanced population of excited states around $\omega_0$, which is tuned to the position of the absorption peak. Such a nonthermal population of states around $\omega_0$ naturally leads to the appearance of the negative weight of the absorption peak, as shown in Fig.~\ref{fig1}(c). In addition to the negative weight, we find a pronounced structure around $\omega=2$ in Fig.~\ref{fig1}(c) as compared to the case of small $t_d$ [Fig.~\ref{fig1}(b)] where an additional structure around $\omega=1$ exists. With increasing $A_0$ up to $0.3$ in Fig.~\ref{fig1}(c), the structure around $\omega=2$ largely diminishes as it is accompanied by the vanishing of the induced charge correlations (not shown here). Since the pumping pulses with $t_d=0.5$ used in Fig.~\ref{fig1}(b) also induce an enhancement of charge correlation but with less extent than that in Fig.~\ref{fig1}(c)~\cite{Lu2012}, the pronounced structure around $\omega=2$ in Fig.~\ref{fig1}(c) may be attributed to a signature of enhanced charge correlation. In fact, the persistence of such an in-gap state is similar to the case of the CDW phase, as will be discussed below.

\begin{figure}
\includegraphics[width=0.45\textwidth]{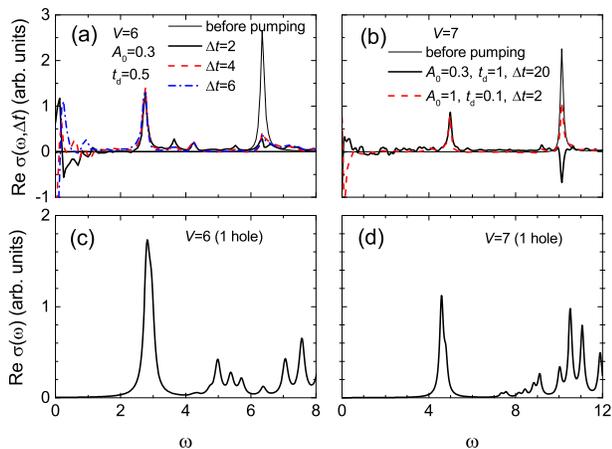}
\caption{(Color online)
$\text{Re}\,\sigma(\omega,\Delta t)$ for the CDW phase of the extended Hubbard model with $U=10$. (a) $\Delta t$ dependence for $L=14$ and $V=6$. $A_0=0.3$ and $t_d=0.5$. (b) Different parameter sets for $L=14$ and $V=7$. $A_0=0.3 (1)$, $t_d=1 (0.1)$, and $\Delta t=20 (2)$ for the solid (dashed) line. The in-gap peak with positive weight at $\omega=5$ is insensitive to the parameter sets. (c) Equilibrium optical conductivity $\text{Re}\,\sigma(\omega)$ for the same parameter set as (a) but away from half filling (seven up-spin electrons and six down-spin electrons). (d) The same as (c) but the same parameter set as (b).
}
\label{fig3}
\end{figure}

\subsection{In-gap states in CDW}

The CDW state is stable for $V>5$. Figure~\ref{fig3}(a) shows the $\Delta t$ dependence of $\text{Re}\,\sigma(\omega,\Delta t)$ for $V=6$. The weight of the absorption peak at $\omega = 6.2$ substantially decreases with pumping. A low-energy structure with both positive and negative weights around $\omega=0.3$ emerges and exhibits a strong $\Delta t$ dependence, which has not been seen in Fig.~\ref{fig1}. Because of a strong time dependence, we consider that the low-energy excitations are not a signature of photoinduced states. On the other hand, a strong in-gap peak appears at $\omega=2.7$ with weak $\Delta t$ dependence. We regard this as a relevant photoinduced state. The $\text{Re}\,\sigma(\omega,\Delta t)$ for $V=7$ also shows a similar in-gap excitation that is independent of the pumping condition and $\Delta t$, as shown in Fig.~\ref{fig3}(b). Therefore, the in-gap state is expected as a characteristic of the CDW phase.

We now discuss the origin of the in-gap state in the CDW phase. Again, we should notice that the pumping creates photo-carriers in the system. However, in the CDW phase, a metallic or low-energy excitation is unstable as seen in Fig.~\ref{fig3}. This is expected to be related to the charge distribution of the CDW ground state with the characteristic arrangement of alternating doubly occupied and empty sites. The photo-carriers are created by transforming pairs of doubly occupied and empty sites to singly occupied ones, but the movement of electrons is still blocked by the presence of neighboring doubly occupied sites. This may lead to a finite-energy excitation inside the original CDW gap. This is confirmed by calculating the optical conductivity for the extended Hubbard model away from half filling with a single hole. Figures~\ref{fig3}(c) and \ref{fig3}(d) exhibit the equilibrium $\text{Re}\,\sigma(\omega)$ for $U=6$ and $U=7$ in the CDW phase, respectively, with seven up-spin electrons and six down-spin electrons in the $L=14$ lattice. In-gap peaks appear at $\omega=2.8$ and $4.6$ for $U=6$ and $7$, respectively. Their energies are close to those of the photoinduced in-gap excitations in Figs.~\ref{fig3}(a) and \ref{fig3}(b), indicating that the presence of carriers is indeed the origin of the in-gap states.

\begin{figure}
\includegraphics[width=0.4\textwidth]{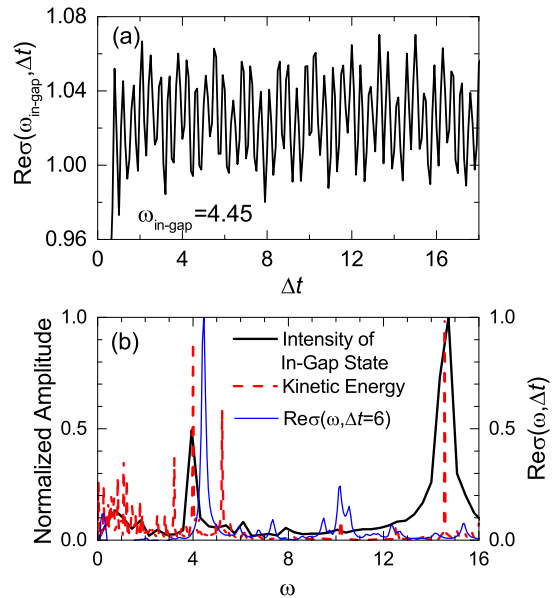}
\caption{(Color online)
(a) $\Delta t$ dependence of intensity of an in-gap state at $\omega_\text{in-gap}=4.45$ in $\text{Re}\,\sigma(\omega,\Delta t)$ for the CDW phase of the $L=10$ extended Hubbard model with $U=10$ and $V=7$. The conditions of the pumping pulse are $A_0=0.3$ and $t_d=0.5$. (b) The amplitude of the Fourier component of the in-gap peak intensity (black solid line) as well as that of kinetic energy (red dashed line). The amplitude is normalized by the maximum value. The blue thin line represents $\text{Re}\,\sigma(\omega,\Delta t=6)$.
}
\label{fig4}
\end{figure}

The intensity of the in-gap state at $\omega_\text{in-gap}=4.45$ for the $L=10$ lattice with $U=10$ and $V=7$ oscillates with $\Delta t$ as shown in Fig.~\ref{fig4}(a). After Fourier analysis, we find characteristic frequencies at $\omega=4$ and $14$ in Fig.~\ref{fig4}(b) (black solid line). The former is very close to the energy of the in-gap state seen in $\text{Re}\,\sigma(\omega,\Delta t=6)$. We find that not only the intensity of the in-gap state but also the kinetic energy, which is an average value of the kinetic terms in $H$ (\ref{H}), displays a pronounced weight in the Fourier spectrum around $\omega_\text{in-gap}$, as seen in Fig.~\ref{fig4}(b) (red dashed line). Similar to the case of SDW, this characteristic oscillation related to the in-gap state is also recognized from the analysis of the remaining main absorption peak (not shown here). This means that the energy of the in-gap state may be detected by measuring the oscillations of the physical quantities. The localization effect of the CDW background on the photoexcited charge carriers might be able to be identified as well in higher-dimensional systems by investigating the superimposed oscillations of time-resolved signals~\cite{Dean2011}. The response at $\omega =14$ is due to high-energy excitations that are due to symmetry restrictions not picked up by the optical conductivity measurement. These excitations are determined by $U$ and $V$, leading to oscillations with a short period, as seen in Fig.~\ref{fig4}(a).

\section{Conclusions}

In summary, we have analyzed the physical properties of photoinduced excitations generated by a short laser pulse and their influence on the time evolution of various physical properties in the 1D extended Hubbard model at half filling. We have calculated the time-dependent optical conductivity driven by a transient laser pulse and clarified the difference in the nonequilibrium optical response between the SDW and the CDW phases. The photoinduced in-gap excitations exhibit a qualitatively different structure in the SDW or CDW ground state. In the SDW, a dominant in-gap excitation emerges at the low-energy region, reflecting the energy difference of the even and odd parity of photoexcited states, where the even- (odd-) parity state corresponds to the optically forbidden (allowed) state. In the CDW, an in-gap state appears in the middle of the Mott gap and its origin can be attributed to photogenerated carriers. The signature of the in-gap excitations in both cases emerges in the time evolution of physical quantities as a characteristic oscillation with a characteristic frequency roughly matching the energy of the in-gap excitations. Such a phenomenon has been reported in the time-dependent optical reflectivity measurement on ET-F$_2$TCNQ, a prototype 1D Mott insulator~\cite{Wall2011}. Observation of this characteristic frequency represents an experimental probe to identify  photoinduced in-gap states.

\begin{acknowledgments}
H.L. acknowledges support from the National Natural Science Foundation of China (NSFC) (Grants No. 11325417, and No. 11474136), and the open project of State Key Laboratory of Theoretical Physics (SKLTP), ITP, CAS.
T.T. acknowledges support by the Grant-in-Aid for Scientific Research (26287079) from
MEXT and the Strategic Programs for Innovative Research (SPIRE) (hp140215), the Computational Materials Science Initiative (CMSI). 
J.B. acknowledges support by Program No. P1-0044 of the Slovenian Research Agency, and the Center for Integrated Nanotechnologies, a U.S. Department of Energy Office of Science User Facility.
A part of the numerical calculations was performed in the supercomputing facilities in ISSP in the University of Tokyo.
\end{acknowledgments}

\nocite{*}


\end{document}